\documentstyle[aps,multicol,epsfig,eqsecnum]{revtex}

\title{Quantum interference with molecules: The role of internal states}
\author{Mark Hillery$^{1,2}$, Leonard Mlodinow$^{1}$\footnote{Permanent address: Alexei Nicolai Inc., 1970
La France Avenue, South Pasadena, CA 91030}, and Vladim\'{\i}r Bu\v{z}ek$^{2,3}$}
\address{
$^{1}$~Department of
Physics, Hunter College of CUNY, 695 Park Avenue,  New York, NY 10021 USA\\
$^{2}$~Research Center for Quantum Information,  Slovak
Academy of Sciences, 845 11 Bratislava, Slovakia\\
$^{3}$~{\em Quniverse}, L{\'\i}\v{s}\v{c}ie \'{u}dolie 116, 841 04 Bratislava, Slovakia
}
\begin{document}
\maketitle
\begin{abstract}
Recent experiments have shown that fullerene and fluorofullerene molecules
can produce interference patterns.  These molecules have both rotational
and vibrational degrees of freedom.  This leads one to ask whether these
internal motions can play a role in degrading the interference pattern.
We study this by means of a simple model.  Our molecule consists of two
masses a fixed distance apart.  It scatters from a potential with two
or several peaks, thereby mimicking two or several slit interference.
We find that in some parameter regimes the entanglement between the internal
states and the translational degrees of freedom produced by the potential
can decrease the visibility of the interference pattern.  In particular,
different internal
states correspond to different outgoing wave vectors, so that if several
internal states are excited, the total interference pattern will be the
sum of a number of patterns, each with a different periodicity.  The overall
pattern is consequently smeared out.  In the case of two different peaks,
the scattering from the different peaks will excite different internal states
so that the path the molecule takes become entangled with its internal state.
This will also lead to degradation of the interference pattern.  How these
mechanisms might lead to the emergence of classical behavior is discussed.
\end{abstract}

%\vspace{2cm}
%\pagebreak

%\begin{multicols}{2}
\section{Introduction}
How big does an object have to be before quantum mechanical effects
disappear?  A cat is presumably too large, but a molecule may not be.
Interference with ${\rm H}_{2}$ molecules was demonstrated as early as 1930
\cite{stern}, and there have been a number of subsequent experiments
with small molecules \cite{borde,lisdat,chapman,clauser,grisenti}.
In a series of recent experiments, the group at the University of Vienna
has shown that molecules consisting of many atoms, such as fullerenes or
fluorofullerenes can produce an interference pattern after travelling
through a grating \cite{zeilinger1,zeilinger2}.  Molecules, however, come
in many sizes, so one might expect that, in regard to their translational
degrees of freedom, small molecules behave quantum mechanically while large
ones do not.  Where is the boundary between these behaviors, and what causes
the transition, if indeed it takes place?

There are several mechanisms that can destroy an interference pattern produced by
matter waves. For example,
a decoherence of molecule in a beam can be caused by to collisions with lighter particles from the environment
(see e.g. Refs.~\cite{hornberger2003,hornberger2004}). Other destructive environmental influences are
grating vibrations, the finite size of a grating, or even  Coriolis forces (see e.g. Ref.~\cite{stibor2005}). Obviously, the chief destroyer
of quantum coherences is thermal radiation through which the molecules become entangled with external (light) degrees of freedom.
This mechanism, which leads to  disappearance of the interference pattern, has been studied theoretically by several groups.
The interference pattern found by the group in Vienna
disappears if the internal temperature of the molecules is sufficiently
high (2000 K)\cite{zeilinger3}. This experimental result has been analyzed in detail  and theoretically
explained by K. Hornberger {\it et al.} \cite{hornberger2004,hornberger2005}. In particular, these authors have studied
the effect of thermal radiation at different temperatures of the molecule on the decoherence of fullerenes, taking into account that
these molecules are not black body radiators. Under this assumption they have been able to find a good agreement with experimental results
\footnote{In other studies (see e.g. Refs.~\cite{saverio,alicki2002})
 an assumption has been made, that the molecule is a black-body radiator and consequently
these results, while they illustrate the effect of this type of decoherence, do not apply directly to experiments in Vienna.}.

Another possible source of decoherence is the coupling between the
translational motion of the molecule and its internal states \cite{hornberger2004}.  The
vibrational and rotational states of the molecule can be thought of as
a reservoir that the molecule carries with it.  When the molecule passes
through a region in which there is a potential, the translational motion of
the molecule and its internal modes can become entangled.  Therefore, the
molecule's internal reservoir can cause different outgoing momenta or
different paths the molecule can take through the region of nonzero
potential to decohere.

Here we wish to examine the effect of the internal states of the molecule
on its interference.  We shall do so by considering a simple model that is
a version of two-slit interference, the most fundamental quantum mechanical
interference phenomenon.  Our molecule will be a rigid rotator, consisting of
two equal masses separated by a fixed distance, and, for simplicity, it
will be confined to two dimensions.  The molecule has both translational and
rotational degrees of freedom.  It will scatter off of a potential,
which will initially be taken to be one consisting
of two peaks (see Fig.~1).  The molecule will interact with both parts of the potential,
and the two scattered waves will both be incident on a detector at some
distant point.  The result of this process is described by the scattering
cross section, and we shall be interested in whether or not it exhibits
interference fringes.  It is straightforward to extend our model to the case
in which the potential has more than two peaks, in particular to the experimentally
relevant situation in which it represents a diffraction grating, and interference
pattern resulting from the scattering from this type of potential will be
studied as well.

Entanglement between internal and translational states can, in fact,
contribute to washing out an interference pattern, but only in certain
parameter regimes. We shall examine two ways in which this can happen.  The
first is a result of the fact that after the scattering, different internal
states of the molecule have different wave vectors describing their
translational motion, i.e.\ the outgoing momenta and the internal states of
the molecule become entangled.  For example, if the molecule is initially
in a plane wave state of
its translational motion and not rotating, and the outgoing state
contains components that correspond to rotational motion of the molecule,
then these components will have a wave vector whose magnitude is smaller
than that of the original wave vector.  This follows from energy conservation;
some of the translational kinetic energy has been transformed into
rotational kinetic energy by the scattering.  This mechanism was proposed
by Hegerfeldt and K\"{o}hler as a means of separating an
excited state of a molecule from its ground state \cite{hegerfeldt}.
Molecules, which either have or have not been excited,
pass through a transmission grating, and, because of the difference
in their wave vectors, the ground and excited state molecules scatter in
different directions.  In our case, the different parts of the state of
the scattered molecule, each part having a different wave vector, result
in a total interference pattern made up of patterns with different spacings
between their peaks, and this causes a smearing out of the overall pattern.
The second way in which the pattern can be degraded comes into play if the
potential is different in different regions, in particular if it consists
of two peaks, which are not the same. Then the path the molecule takes and
its internal state can become entangled.  If the internal states are
sufficiently different, then they will reveal which path the molecule took,
and there will be no interference pattern.  Our model will allow us to study
both of these mechanisms.

We calculate the cross section for our scattering in the Born approximation,
which we must modify to take into account the internal (rotational)
states of the molecule.  This calculation is relatively standard, so we
have relegated it to an appendix.  We begin the next section with the
expression for the cross section, and proceed to analyze its implications.
\begin{figure}
\label{fig1}
\begin{center}
\includegraphics[width=8cm]{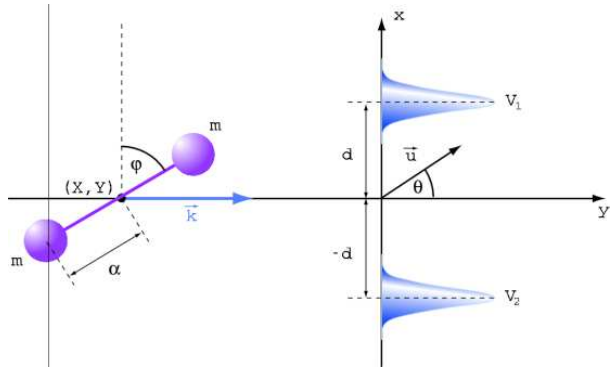}
\end{center}
\caption{
Schematic representation of a scattering of a two-atom molecule on a potential composed
of two (Gaussian) peaks. The two ``atoms'' of the molecule have a mass $m$, and each located a distance $\alpha$ from
the center of mass of the molecule.  The location of the center of mass
is given by $(X,Y)$. The angle
between the line connecting the two masses and the positive $x$ axis is
$\phi$, and the angular momentum corresponding to the rotation of the
molecule in the $x$-$y$ plane is $L$. The angle between the wave vector of
the outgoing molecule and the positive $y$ axis is denoted by $\theta$,
and $\hat{\mathbf{u}}$ is a unit vector pointing in the direction
of the outgoing wave vector.
}
\end{figure}

\section{Cross section}
We begin by describing our system in more detail.  The molecule consists
of two ``atoms'', both of mass $m$, and each located a distance $\alpha$ from
the center of mass of the molecule.  The location of the center of mass
is given by $(X,Y)$ and its momentum by $(P_{x},P_{y})$.  The angle
between the line connecting the two masses and the positive $x$ axis is
$\phi$, and the angular momentum corresponding to the rotation of the
molecule in the $x$-$y$ plane is $L$.  There is also a potential with
which the particles interact, $V$. In Fig.~1 we present a schematic
picture of the physical situation we consider.
The Hamiltonian describing the
system is
\begin{equation}
\label{ham}
H=\frac{1}{4m}(P_{x}^{2}+P_{y}^{2})+\frac{1}{2I}L^{2}+V(X+\alpha\cos\phi ,
Y+\alpha\sin\phi )+V(X-\alpha\cos\phi ,Y-\alpha\sin\phi ) ,
\end{equation}
where $I=2m\alpha^{2}$ is the moment of inertia of the molecule.  We shall
label the states of the system, $|{\mathbf k},l\rangle$ by the wave vector
of the center of mass, ${\mathbf k}$, and the quantum number of the rotational
motion, $l$.  The states are normalized so that
\begin{equation}
\langle{\mathbf k},l|{\mathbf k}^{\prime},l^{\prime}\rangle =\delta^{2}
({\mathbf k}-{\mathbf k}^{\prime})\delta_{l,l^{\prime}}\, .
\end{equation}

Now we consider the situation in which the molecule comes in from the negative
$y$ direction with an initial state
\begin{equation}
|\Psi_{in}\rangle = \sum_{l=-\infty}^{\infty}\psi_{l}|{\mathbf k}=k
\hat{\mathbf{y}},l\rangle .
\end{equation}
Let $\theta$ be the angle between the wave vector of the outgoing molecule
(the wave vector after the scattering has take place) and the positive
$y$ axis, and $\hat{\mathbf{u}}$ be a unit vector pointing in the direction
of the outgoing wave vector.  The cross section for the scattering is given by
\begin{eqnarray}
\label{gencrosssec}
\sigma (\theta ) =  (2\pi )^{3}\frac{4m^{2}}{k}
\sum_{l,l^{\prime}=-\infty}^{\infty}\Theta \left(\frac{k^{2}}{4m}+
\frac{l^{2}}{2I}-\frac{(l^{\prime})^{2}}{2I}\right) |\psi_{l}|^{2}
%\nonumber \\
% & &
 |\langle \kappa (k,l,;l^{\prime})\hat{\mathbf{u}},l^{\prime}|V|
k\hat{\mathbf{y}},l\rangle |^{2} ,
\end{eqnarray}
where $\Theta (x) =1$ if $x\geq 0$ and $\Theta (x)=0$, if $x<0$, and
\begin{equation}
\label{kappa}
\kappa (k,l;l^{\prime}) = 2\sqrt{m}\left(\frac{k^{2}}{4m}+\frac{l^{2}}{2I}
-\frac{(l^{\prime})^{2}}{2I}\right)^{1/2} .
\end{equation}
This is to be compared with the cross section for the scattering, in two
dimensions, of a particle of mass M and no internal structure from a
potential $V({\mathbf r})$, which is, in the Born approximation,
\begin{equation}
\label{noint}
\sigma (\theta )=\frac{M^2}{2\pi k}\left| \int d^{2}r
e^{-ik(\hat{\mathbf{u}}-\hat{\mathbf{y}})\cdot {\mathbf r})}
V({\mathbf r})\right|^{2} .
\end{equation}

We now need to calculate the matrix element of the potential.  First we
shall assume that the potential is the sum of two terms, $V({\mathbf r})
= V_{1}({\mathbf r})+ V_{2}({\mathbf r})$, where $V_{1}({\mathbf r})$ is
centered about the point $(d,0)$ and $V_{2}({\mathbf r})$ is centered
about the point $(-d,0)$.  The distance $2d$ plays the role of the distance
between the slits in a two-slit interference experiment.  The matrix
element appearing in the cross section is then the sum of four terms, as
both $V_{1}$ and $V_{2}$ are evaluated at the points $(X+\alpha\cos\phi ,
Y+\alpha\sin\phi )$ and $(X-\alpha\cos\phi ,Y-\alpha\sin\phi )$, i.e.\ the
coordinates of the two masses of the molecule.  We define, for $j=1,2$,
\begin{equation}
\tilde{V}_{j}(k_{x},k_{y})=\frac{1}{2\pi}\int dx\int dy e^{-i(k_{x}x+k_{y}y)}
V_{j}(x,y)  .
\end{equation}
For the matrix element of $V_{j}(X+\alpha\cos\phi ,Y+\alpha\sin\phi )$, which
we shall denote by $V_{j}^{(+)}$, we then have
\begin{eqnarray}
\langle \kappa\hat{\mathbf{u}},l^{\prime}|V_{j}^{(+)}|k\hat{\mathbf{y}},
l\rangle
 & = & \frac{1}{(2\pi)^{3}}\int_{-\infty}^{\infty} dX \int_{-\infty}^{\infty}
dY \int_{0}^{2\pi} d\phi e^{i(\kappa\hat{\mathbf{u}}-k\hat{\mathbf{y}})
\cdot {\mathbf R}}  e^{i(l-l^{\prime})\phi}V_{j}(X+\alpha\cos\phi ,Y+\alpha\sin\phi )
\nonumber \\
 & = & \frac{1}{2\pi}\tilde{V}_{j}(-\kappa\sin\theta ,k-\kappa\cos\theta )
 \int_{0}^{2\pi} d\phi e^{i(l-l^{\prime})\phi}
 \exp[{-i\kappa\alpha\sin\theta\cos\phi}] \exp[{i(k-\kappa\cos\theta )\alpha
\sin\phi }] \nonumber \\
 & = & \frac{1}{(2\pi )^{1/2}}e^{-i(l-l^{\prime})\mu}J_{l-l^{\prime}}
(\alpha |\kappa\hat{\mathbf{u}}-k\hat{\mathbf{y}}|)\tilde{V}_{j}
(k\hat{\mathbf{y}}-\kappa\hat{\mathbf{u}}) .
\end{eqnarray}
The angle $\mu$ satisfies
\begin{equation}
\label{sincosmu}
\sin\mu = \frac{\kappa\sin\theta}{|\kappa\hat{\mathbf{u}}-k\hat{\mathbf{y}}|}\; ;
\hspace{1cm} \cos\mu
= \frac{\kappa\cos\theta -k}{|\kappa\hat{\mathbf{u}}-k\hat{\mathbf{y}}|} ,
\end{equation}
so that
\begin{equation}
\mu = \tan^{-1}\left[ \frac{\kappa\sin\theta}{\kappa\cos\theta -k}\right] .
\end{equation}
Note that the quadrant in which $\mu$ lies is specified by
Eq.\ (\ref{sincosmu}).
The evaluation of the matrix element of $V_{j}(X-\alpha\cos\phi ,Y-\alpha
\sin\phi )$, which we denote by $V_{j}^{(-)}$ is similar, and we find
\begin{equation}
\langle \kappa\hat{\mathbf{u}},l^{\prime}|V_{j}^{(-)}|k\hat{\mathbf{y}},
l\rangle = \frac{1}{(2\pi )^{1/2}}
e^{-i(l-l^{\prime})\mu}(-1)^{l-l^{\prime}}J_{l-l^{\prime}}
(\alpha |\kappa\hat{\mathbf{u}}-k\hat{\mathbf{y}}|)\tilde{V}_{j}
(k\hat{\mathbf{y}}-\kappa\hat{\mathbf{u}}) .
\end{equation}
Consequently, we have that
\begin{equation}
\langle \kappa\hat{\mathbf{u}},l^{\prime}|V|k\hat{\mathbf{y}},l\rangle =
\frac{1}{2\pi} e^{-i(l-l^{\prime})\mu}[1+(-1)^{l-l^{\prime}}]J_{l-l^{\prime}}
(\alpha |\kappa\hat{\mathbf{u}}-k\hat{\mathbf{y}}|)\sum_{j=1}^{2}
\tilde{V}_{j}(k\hat{\mathbf{y}}-\kappa\hat{\mathbf{u}}) .
\end{equation}

Let us begin by considering the case in which $V_{1}$ and $V_{2}$ are both
Gaussians with the same width
\begin{eqnarray}
V_{1}({\mathbf r})& = & V_{0}e^{-|{\mathbf r}-d\hat{\mathbf{x}}|^{2}/\Delta^{2}}\; ;
\nonumber \\
V_{2}({\mathbf r}) & = & V_{0}
e^{-|{\mathbf r}+d\hat{\mathbf{x}}|^{2}/\Delta^{2}} .
\end{eqnarray}
With this potential, the cross section becomes
\begin{eqnarray}
\label{gausscrosssec}
\sigma (\theta ) & = & \frac{8\pi m^{2}\Delta^{4}V_{0}^{2}}{k}
\sum_{l,l^{\prime}=-\infty}^{\infty} \Theta \left(\frac{k^{2}}{4m}+
\frac{l^{2}}{2I}-\frac{(l^{\prime})^{2}}{2I}\right)
[1+(-1)^{l-l^{\prime}}]^{2} \nonumber \\
&\times & e^{-\Delta^{2}|\kappa\hat{\mathbf{u}}-k\hat{\mathbf{y}}|^{2}/2}
J_{l-l^{\prime}}^{2}(\alpha |\kappa\hat{\mathbf{u}}-k\hat{\mathbf{y}}|)
\cos^{2}(\kappa d\sin\theta ) |\psi_{l}|^{2} .
\end{eqnarray}
The presence of an interference pattern can be seen from the presence of
the cosine term.

We shall examine this expression in a number of parameter regions.  First,
consider the case in which $\Delta \gg \alpha$, i.e.\ the potential varies
slowly over distances of the order of the size of the molecule.  We would
not expect the fact that the molecule has internal degrees of freedom to
play much of a role in this case, and this is, indeed what we find.  Note
that, because of the Gaussian factor, the cross section will be small unless
$|\kappa\hat{\mathbf{u}}-k\hat{\mathbf{y}}|\Delta$ is of order one or smaller.
We have that
\begin{eqnarray}
|\kappa\hat{\mathbf{u}}-k\hat{\mathbf{y}}|\Delta & > & |\kappa -k|\Delta
\nonumber \\
 & > & \left| \left[ (k\Delta )^{2}+\left(\frac{\Delta}{\alpha}\right)^{2}
[l^{2}-(l^{\prime})^{2}] \right]^{1/2}-k\Delta\right| .
\end{eqnarray}
From this expression we see that for $\Delta \gg \alpha$, we will have that
$|\kappa\hat{\mathbf{u}}-k\hat{\mathbf{y}}|\Delta \gg 1$ unless
$l=l^{\prime}$, and this implies that only the terms with $l=l^{\prime}$ will
contribute to the sum in Eq.\ (\ref{gausscrosssec}).  In that case we have
that $\kappa = k$, and
\begin{eqnarray}
\sigma (\theta )  =  \frac{32\pi m^{2}\Delta^{4}V_{0}^{2}}{k}
e^{-(k\Delta )^{2}|\hat{\mathbf{u}}-\hat{\mathbf{y}}|^{2}/2}
 J_{0}^{2}(k\alpha |\hat{\mathbf{u}}-\hat{\mathbf{y}}|)
\cos^{2}(kd\sin\theta )  .
\end{eqnarray}
Note that, as expected, all of the dependence on the internal states is
gone.  If, in addition, the molecule is much smaller than the wavelength
corresponding to its center-of-mass motion, then the Bessel function can be
replaced by $1$, and all vestige of internal structure disappears.  In fact,
in this limit the scattering cross section becomes the same as that of
a particle with no internal structure and a mass of $2m$ scattering from
a potential $2(V_{1}({\mathbf r})+V_{2}({\mathbf r}))$ (see Eq.\ (\ref{noint})).

In the experiments with fullerenes, the ratio of the slit separation to
the de Broglie wavelength of the molecules was of the order of $10^4$,
and the scattering only at angles near the forward direction was observed.
Let us see what our toy model predicts for a situation similar to this
one.  We shall assume that $kd\sim 10^4$, $\alpha < d$ but the two are of
the same order of magnitude, and, for simplicity, that our molecule starts
in the $l=0$ state.  In addition, we shall concentrate our attention on
angles near the forward direction, in particular, $\theta\sim 10^{-4}$.
We now have that
\begin{equation}
\kappa = k\left[ 1-\frac{(l^{\prime})^{2}}{(k\alpha )^{2}}\right]^{1/2} ,
\end{equation}
so that the only angular momentum states that can be excited are the ones
satisfying $k\alpha \geq l^{\prime}$.  Since $k\alpha \sim 10^4$, it seems
that many internal states can be excited, and that this could have an
effect on the interference pattern.  In particular, the interference
pattern is given by the $\cos (\kappa d\sin\theta )$ factor, so that if
many values of $l^{\prime}$, and hence many values of $\kappa$, are
allowed, then the interference pattern could be smeared out.

This does not happen in this situation.  As was noted earlier, in order for
the Gaussian factor not to cut everything off, we must have $|k-\kappa |$
of order one or less, that is
\begin{equation}
k\Delta \left\{ 1-\left[ 1-\frac{(l^{\prime})^{2}}{(k\alpha )^{2}}
\right]^{1/2}\right\} <1  .
\end{equation}

\begin{figure}
\label{fig2}
\begin{center}
\includegraphics[width=13cm]{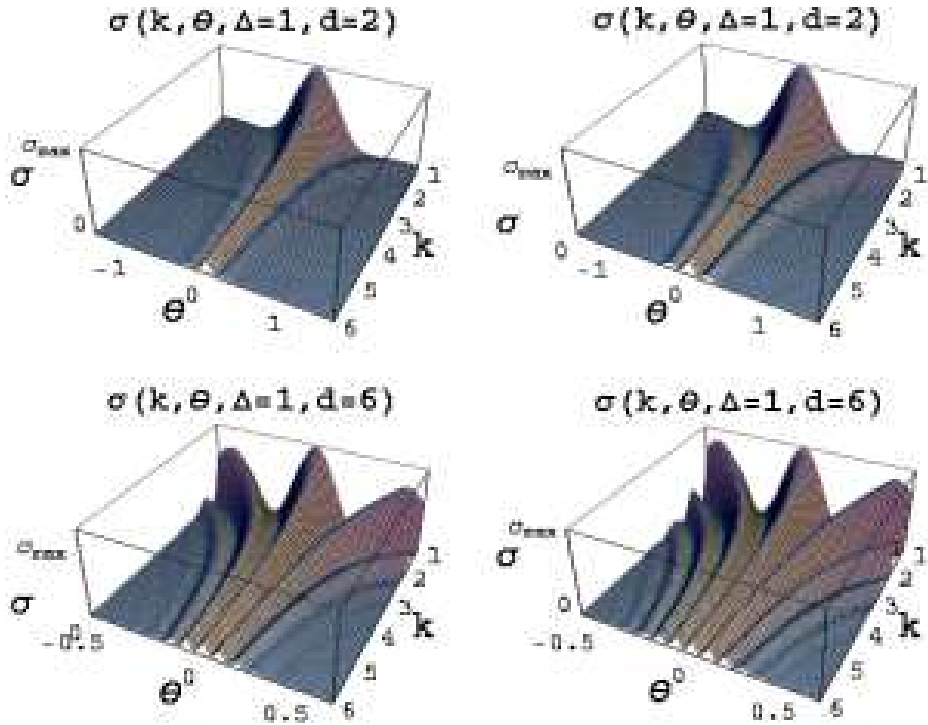}
\end{center}
\caption{
We plot the cross section $\sigma(\theta)$ as a function of $\theta$ and the momentum $k$.
We compare two cases: a molecule with (left column) and without (right column)
internal degrees of freedom. From the figure it is obvious that the internal structure
of the molecule causes a suppression of the interference pattern. The cross section $\sigma$ in the presence of
the internal structure of the molecule is represented by Eq.(2.14) where we assume that the incident molecule is in the
state with $l=0$. The cross section corresponding to the case with a molecule that does not have an internal structure
is given by Eq.(2.21).  Otherwise the molecules are considered to be the same, i.e. we assume units such that $m=1$ and $\hbar=1$.
The two Gaussian peaks are characterized by $V_0=1$ and $\Delta=1$. The distance between each scattering peak and the origin of the coordinate
system is $d=2$ (the upper row) and $d=6$ (the lower row). We observe that with larger
$d$ the frequency of interference oscillations is larger.
The suppression of oscillations due to the presence of internal molecular states is clearly
seen in both cases, i.e. $d=2$ and $d=6$.
}
\end{figure}
Let us assume that $\Delta < d$, but that $k\Delta \gg 1$.  Then in order
to satisfy the above condition, we must have $l^{\prime}/(k\alpha )\ll 1$,
and, in particular, $l^{\prime}/(k\alpha ) < 1/\sqrt{k\Delta}$.  This
implies that the values of $\kappa$ that contribute to the sum in the
expression for the cross section, lie in the range between $k$ and
$k-(1/\Delta )$, that is, all of these values are very close to $k$.
Making the approximation that $\kappa = k$, we find that
\begin{equation}
|\kappa\hat{\mathbf{u}}-k\hat{\mathbf{y}}|=\sqrt{2}k(1-\cos\theta )^{1/2} ,
\end{equation}
and
\begin{eqnarray}
\label{2peakexp}
\sigma (\theta ) & = & \frac{8\pi m^{2}\Delta^{4}V_{0}^{2}}{k}
\sum_{l^{\prime}=-\infty}^{\infty} \Theta \left( k^{2}
-\frac{(l^{\prime})^{2}}{\alpha^{2}}\right)[1+(-1)^{l^{\prime}}]^{2}
\nonumber \\
&\times & e^{-(k\Delta)^{2}(1-\cos\theta )/2}
J_{l^{\prime}}^{2}(\sqrt{2}\alpha k(1-\cos\theta )^{1/2} )
\cos^{2}(kd\sin\theta )  .
\end{eqnarray}
In the range $\theta \sim (1/kd) \sim 10^{-4}$, we find that the quantities
$k\Delta (1-\cos\theta )^{1/2}$, $k\alpha (1-\cos\theta )^{1/2}$, and
$kd\sin\theta$ are all of order one, and the cross section exhibits
strong interference fringes.

In our examples so far, the internal states have not influenced the interference pattern.  One
regime in which they do is when $k\Delta$, $k\alpha$, and $kd$ are roughly of order one.
In Fig.~2 we present the cross section $\sigma(\theta)$ as a function of $\theta$ and the momentum $k$ for parameters in this range.
We compare two cases: Firstly, we consider the case when the molecule has an internal structure and the cross section $\sigma(\theta)$ is
described by Eq.~(2.14). Secondly, we analyze the situation when the molecule has no internal structure. In all cases when we are comparing scattering with and without internal structure,
the molecule without internal structure is taken to have a mass of $2m$, and the potential
acting on it is twice that of the potential for the molecule with internal structure (this is
to compensate for the fact that for the molecule with internal structure, the potential acts
on each particle).
For this case using the general expression
given by Eq.~(2.6) we can derive the cross section for the scattering potential with two Gaussian peaks in the form:
\begin{eqnarray}
\sigma (\theta ) = \frac{32\pi m^{2}\Delta^{4}V_{0}^{2}}{k}
\exp\left[-\Delta^{2} k^2(1-\cos\theta )\right]
\cos^{2}(kd\sin\theta )  .
\end{eqnarray}
From Fig.~2 it is obvious that the internal structure
of the molecule causes a suppression of the interference pattern. In the figure we use units and values of the parameters involved in the
expressions for the cross section such that $k\Delta$, $k\alpha$ and $kd$ are not too large (of the order between 1 and 10 in dimensionless
units used in the figure).

The actual experiments use a grating instead of two slits, and this it is
straightforward to incorporate this into our model.  Suppose the the
potential consists of $2N+1$ peaks centered on the $x$ axis, so that
\begin{equation}
V({\mathbf r})=\sum_{n=-N}^{N}v({\mathbf r}-nd\hat{\mathbf{x}}) ,
\end{equation}
where $v({\mathbf r})$ is the potential for one of the peaks.  We then find
that
\begin{eqnarray}
\tilde{V}({\mathbf k}) & = & \sum_{n=-N}^{N}e^{-ik_{x}nd}\tilde{v}({\mathbf k})
 =  \frac{\sin [k_{x}d(2N+1)/2]}{\sin (k_{x}d/2)}\tilde{v}({\mathbf k}) .
\end{eqnarray}
In the case that the peaks are Gaussian,
\begin{equation}
v({\mathbf r})=V_{0}e^{-|{\mathbf r}|^{2}/\Delta^{2}} ,
\end{equation}
the cross section becomes
\begin{eqnarray}
\label{grating}
\sigma (\theta ) & = & \frac{2\pi m^{2}\Delta^{4}V_{0}^{2}}{k}
\sum_{\l^{\prime}=-\infty}^\infty\Theta \left( k^{2}
-\frac{(l^{\prime})^{2}}{\alpha^{2}}\right)[1+(-1)^{l^{\prime}}]^{2}
\nonumber \\
& \times& e^{-(\Delta)^{2}|\kappa\hat{\mathbf{u}}-k\hat{\mathbf{y}}|^{2}/2}
J_{l^{\prime}}^{2}(\sqrt{2}\alpha |\kappa\hat{\mathbf{u}}-k\hat{\mathbf{y}}| )
\frac{\sin^{2}[\kappa d(2N+1)\sin\theta )/2]}{\sin^{2}[(\kappa d\sin\theta )/2]} .
\end{eqnarray}
If  we are in the regime in which the experiments were done, i.e.\ $kd\sim 10^4$
and $\alpha$ and $\Delta$ both less than $d$ but of similar order of magnitude, the
same considerations as those in the derivation of Eq.~(\ref{2peakexp}) apply,
and we find for the cross section
\begin{eqnarray}
\sigma (\theta ) & = & \frac{2\pi m^{2}\Delta^{4}V_{0}^{2}}{k}
\sum_{\l^{\prime}=-\infty}^\infty\Theta \left( k^{2}
-\frac{(l^{\prime})^{2}}{\alpha^{2}}\right)[1+(-1)^{l^{\prime}}]^{2}
\nonumber \\
& \times& e^{-(k\Delta)^{2}(1-\cos\theta )/2}
J_{l^{\prime}}^{2}(\sqrt{2}\alpha k(1-\cos\theta )^{1/2} )
\frac{\sin^{2}[(kd(2N+1)\sin\theta )/2]}{\sin^{2}[(kd\sin\theta )/2]} .
\end{eqnarray}
We note that near $\theta =0$ the spacing between peaks is roughly $2\pi/[kd(2N+1)]$
(the first peak is at $\theta = 0$, the second, which is not as high, is
between $2\pi/[kd(2N+1)]$ and $4\pi/[kd(2N+1)]$).  In the experiment, a
molecule passes through about $100$ slits (this is the beam width divided by the
slit spacing), which gives a value of $N$ of around $50$.  This implies
an angular spacing between peaks in the pattern near $\theta = 0$ of
$10^{-5}$ to $10^{-6}$, which agrees with what was seen.

\begin{figure}
\label{fig3}
\begin{center}
\includegraphics[width=13cm]{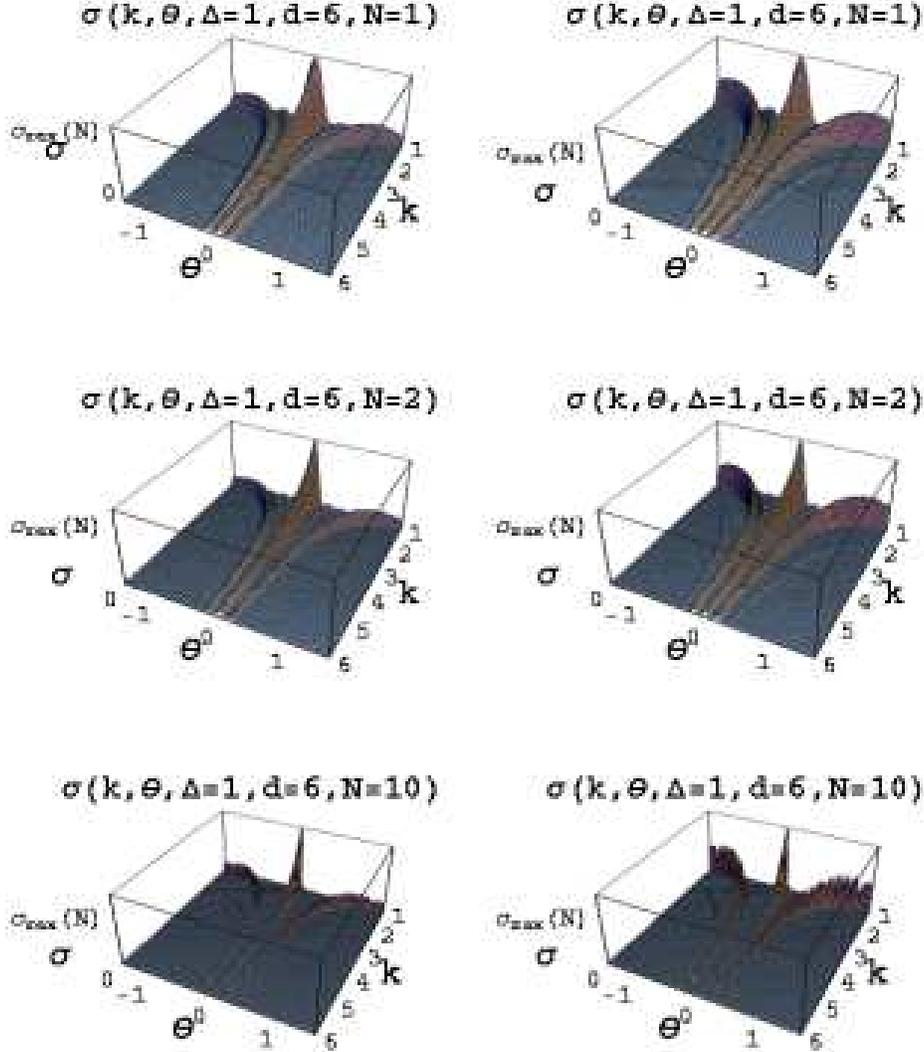}
\end{center}
\caption{
We plot the cross section $\sigma(\theta)$ of the molecule that is scattered by a grating modelled as a potential with $2N+1$ Gaussian peaks.
We present the cross section as a function of $\theta$ and the momentum $k$ for different values of $N$.
We compare two cases: We assume the molecule with (left column) and without (right column)
internal degrees of freedom. From the figure it is obvious that the internal structure
of the molecule causes a suppression of the interference pattern. The cross section $\sigma$ in the presence of
the internal structure of the molecule is represented by Eq.(2.25) where we assume that the incident molecule is in the
state with $l=0$. The cross section corresponding to the case with a molecule that does not have an internal structure
is given by Eq.(2.27).  Otherwise the molecules are considered to be the same and we assume units such that $m=1$ and $\hbar=1$.
The Gaussian peaks are characterized by $V_0=1$ and $\Delta=1$. The distance between adjacent scattering peaks
is $d=6$. We consider three different types of gratings with $N=1$ (i.e. scattering on 3 Gaussian peaks - see the first row),
with $N=2$ (i.e. scattering on 5 Gaussian peaks - see the second row); and with $N=10$ (i.e. scattering on 21 Gaussian peaks - see the third row).
The maximal values of the cross section that are achieved from small $k$ and $\theta=0$ are different for different values of $N$.
Particularly, these values are proportional to $N^2$. In any case, we see a suppression of the
quantum interference patterns due to the presence
of internal molecular states.
}
\end{figure}

In order to investigate the role of internal states for the case of a grating potential,
we again go the the regime in which $k\Delta$, $k\alpha$, and $kd$ are roughly of order one.
The expression for the cross section of a molecule
with no internal structure scattering off of a potential with $2N+1$ identical peaks can be
found by using the general expression given by Eq.~(2.6) and is given by
\begin{eqnarray}
\label{nointgrat}
\sigma (\theta ) = \frac{32\pi m^{2}\Delta^{4}V_{0}^{2}}{k}
\exp\left[-\Delta^{2} k^2(1-\cos\theta )\right]
\frac{\sin^{2}[(kd(2N+1)\sin\theta )/2]}{\sin^{2}[(kd\sin\theta )/2]}.
\end{eqnarray}
We plot the cross sections $\sigma(\theta)$ given by Eqs.~(\ref{grating}) and
(\ref{nointgrat}) in Fig.~3 as a function
of $\theta$ and $k$ for different values of $N$. From the figure we can again conclude
that the presence of the internal structure of the molecule leads to a suppression of the interference pattern.

\section{Different Peaks}
So far we have considered only the case in which both of the peaks in the
potential are the same.  We shall now consider the case in which they
are not.  This will give rise to another mechanism that can decrease
the visibility of the interference pattern.  The different peaks will
give rise to different excitations of the internal states,
and the interference pattern will be proportional to the overlap between
these internal states.  The overlap is related to the information about
the path the molecule followed through the potential.  If the overlap is
zero, then by looking at the internal state of the molecule,
we can determine from which peak it scattered, then there will be no
interference pattern.  If the overlap is greater than zero, then there is
partial information about the path, and the visibility of the interference
pattern is correspondingly reduced.

We shall now assume that the potential is the sum of two terms,
$V({\mathbf r}) = V_{1}({\mathbf r}-d\hat{\mathbf{x}})+
V_{2}({\mathbf r}+d\hat{\mathbf{x}})$, where
\begin{eqnarray}
V_{1}({\mathbf r}) & = & V_{0}\left( 1-\frac{r^{2}}{\Delta^{2}}\right)
e^{-(r/\Delta )^{2}}\; ; \nonumber \\
V_{2}({\mathbf r}) & = & V_{0}e^{-(r/\Delta )^{2}} .
\end{eqnarray}
The Fourier transforms of these potentials are given by
\begin{eqnarray}
\tilde{V}_{1}({\mathbf k})=\frac{1}{8}V_{0}k^{2}\Delta^{4}e^{-(k\Delta)^{2}/4}\; ;
\nonumber \\
\tilde{V}_{2}({\mathbf k})=\frac{1}{2}V_{0}\Delta^{2}e^{-(k\Delta)^{2}/4} ,
\end{eqnarray}
so that the matrix element of the total potential is
\begin{eqnarray}
\langle \kappa\hat{\mathbf{u}},l^{\prime}|V|k\hat{\mathbf{y}},l\rangle & = &
\frac{1}{2\pi}e^{-i(l-l^{\prime})}[1+(-1)^{l-l^{\prime}}]J_{l-l^{\prime}}
(\alpha |\kappa\hat{\mathbf{u}}-k\hat{\mathbf{y}}|) \nonumber \\
 &\times & \frac{1}{2}V_{0}\Delta^{2}e^{-|\kappa\hat{\mathbf{u}}-k
\hat{\mathbf{y}}|^{2}\Delta^{2}/4}\left[ \frac{\Delta^{2}}{4}
|\kappa\hat{\mathbf{u}}-k\hat{\mathbf{y}}|^{2}e^{i\kappa d\sin\theta}
 +e^{i\kappa d\sin\theta}\right] .
\end{eqnarray}

Now consider a particular example.  As usual, we shall assume that the
molecule is initially in the $l=0$ state.  We shall also assume that
$3>k\alpha\geq 1$, which means that the there are three terms in the
$l^{\prime}$ sum (the $l^{\prime}=\pm 1$ terms vanish).  Let us now
examine the terms corresponding to $V_{1}$ and $V_{2}$ in the above
matrix element.  Define the magnitude of the difference between the outgoing
and incoming momenta to be $q= |{\mathbf q}|$ where
${\mathbf q}=\kappa\hat{\mathbf{u}}-k\hat{\mathbf{y}}$.
The $V_{1}$ contribution at $l^{\prime}=0$ is proportional to $J_{0}(\alpha q)
q^{2}\exp [-(q\Delta /2)^{2}]$ while the $V_{2}$ contribution is proportional
to $J_{0}(\alpha q)\exp [-(q\Delta /2)^{2}]$.  The function $q^{2}\exp
[-(q\Delta /2)^{2}]$ has a maximum at $q=2/\Delta$.  If $\alpha$ and
$\Delta$ are chosen so that this maximum occurs at a zero of
$J_{0}(\alpha q)$, then the contribution of $V_{1}$ to the scattering at
$l^{\prime}=0$ will be greatly suppressed.  What will happen then is that
$V_{1}$ will scatter the molecule primarily into the $l^{\prime}=\pm 2$
states and $V_{2}$ will scatter it primarily into the $l^{\prime}=0$
state.  The result will be a suppression of the interference pattern.

Let us make this more specific.  First, the cross-section for this choice
of potential and the molecule initially in the $l=0$ state is given by
\begin{eqnarray}
\sigma (\theta ) & = & \frac{2\pi m^{2}V_{0}^{2}\Delta^{4}}{k}
\sum_{l^{\prime}=-\infty}^{\infty}[1+(-1)^{l^{\prime}}]^{2}J_{l^{\prime}}^{2}
(\alpha q)
e^{-(q\Delta )^{2}/2}\left[ 1 + \frac{(q\Delta )^{4}}{16}
+\frac{(q\Delta )^{2}}{2}\cos (2\kappa d\sin\theta )\right] .
\end{eqnarray}
For the same choice of the potential but for the molecule without an internal structure we find
the cross section using Eq.~(2.6) in a form
\begin{eqnarray}
\sigma (\theta ) & = & \frac{8\pi m^{2}V_{0}^{2}\Delta^{4}}{k}
e^{-(q\Delta )^{2}/2}\left[ 1 + \frac{(q\Delta )^{4}}{16}
+\frac{(q\Delta )^{2}}{2}\cos (2\kappa d\sin\theta )\right] .
\end{eqnarray}
In Fig.~4 we plot the cross sections $\sigma(\theta)$ given by Eqs.~(3.4) and (3.5).
We consider the case when $k\alpha =5/2$, $k\Delta = 3/2$, and $kd=4$.  In this case we
see that the interference pattern is almost completely suppressed by decoherence due to
the internal states of the molecule.
\begin{figure}
\label{fig4}
\begin{center}
\includegraphics[width=6cm]{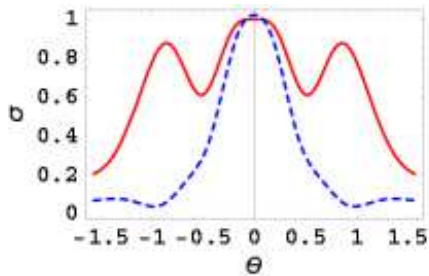}
\end{center}
\caption{
We plot the cross section $\sigma(\theta)$ of the molecule that is scattered by a potential composed of two non-identical peaks.
The potential is described by Eq.~(3.1). We consider two situations: When the molecule has an internal structure
(dashed line) given by Eq.~(3.4)  and when the molecule does not have an internal structure (solid line) given by Eq.~(3.5).
When internal states of the molecule are present, then
depending on which path the molecule follows through the potential, that is which peak it
scatters from,
different (distinguishable) internal states of the molecule are excited. Consequently, in principle the path  can be determined, and the interference
is suppressed (this situation situation is very similar to that discussed in papers on ``which-way information'' in double-slit
experiments - see e.g. Refs.~[18,19]).
%(see e.g. \cite{englert1996,durr1998}).
Here the cross section is given by Eq.~(3.4). We consider parameters such that $k\alpha =5/2$, $k\Delta = 3/2$, and $kd=4$,
and we obtain the cross section given by the dashed line.
When the molecule has no internal structure, the path cannot be determined and the interference pattern is preserved (solid line).
The cross section $\sigma(\theta)$ in this case is given by Eq.~(3.5) and for the same
values of parameters as before (i.e. $k\alpha =5/2$, $k\Delta = 3/2$, and $kd=4$).
}
\end{figure}

\section{Conclusion}
The coupling of internal and translational degrees of freedom of an object
can lead to the degradation of an interference pattern produced by the
scattering of the object from a potential.  This is a result of the entanglement
between the internal and translational degrees of freedom of the object produced by
the potential.  We have studied two different forms this entanglement can take.
In the first different outgoing wave vectors become entangled with different internal
states producing an interference pattern made up of parts with different
periodicities, which leads to a smearing the overall pattern.  The second
results from the entanglement of the internal states and different paths the molecule
can follow through the potential.

Even though our model was very simple, it is possible to draw some conclusions
from the results based on it, and to put forward some conjectures.  In order
for internal states to be excited, they must be of sufficiently low energy.
If the molecule is of size L and mass M, the lowest rotational mode will have
an energy of order $1/ML^{2}$.  We would expect the mass of the molecule to
scale roughly as its volume, which means that $M\sim L^{3}$, so that
the energy of a low-lying rotational state is proportional to $1/L^{5}$.
There will also be vibrational modes.  The energy of the low-lying acoustic
phonon modes will be proportional to $1/L$.  In both cases, it is clear that
the larger the object, the more low energy internal states it will have.

In our simple model, the coupling between the translational motion and the
internal state with angular momentum $l$ (if the molecule was initially in
the zero angular momentum state) is determined by the product
$J_{l}(\alpha q)\tilde{V}({\mathbf q})$.  Therefore, if the Fourier transform
of the potential is significant in the region near $l/\alpha$, where
$J_{l}(\alpha q)$ is largest in absolute value, then there will be a
a good chance of exciting the
$l$th rotational state.  This implies that in order to excite internal
states, the potential must vary on a scale of the order of the size of the
object or smaller.  Clearly this restriction is much less stringent for a
large object than a small one.

These considerations suggest that it is easier to excite internal states in
a large object than in a small one.  There are more low lying states, and
for a given potential, the probability of exciting one of these states
is greater for the larger object, because the conditions for doing so are
less restrictive.  This further suggests that it is more
likely for the translational motion of a large object to become entangled
with its internal states than would be the case of a smaller object.

This gives us a possible mechanism for the emergence of classical behavior
for an object with internal structure moving in a potential.  If we consider
the path-integral description of its dynamics, each of the paths it can
follow will correspond to different internal excitations.  Therefore, these
paths will decohere, and instead of a coherent superposition of paths, we
will have an incoherent one.  In addition, the classical path will be the
most probable, so that the object will simply follow this path through
the potential.

There are clearly many gaps to be filled in before we can claim that this
picture is correct.  The calculations in this paper are a first step.
More sophisticated models and treatments are called for.  However, even
this very simple model shows that internal states can act as a reservoir
and cause different states of translational motion to decohere.

\acknowledgements This research was supported by the National
Science Foundation under grant PHY 0139692. In addition this work was
supported in part by the European Union  projects QGATES and
CONQUEST,  by the Slovak Academy of Sciences via the project CE-PI
and by the project APVT-99-012304. We thank Klaus Hornberger and Markus Arndt for very
helpful correspondence.

\appendix
\section*{Derivation of the cross section}
Here we calculate the cross section for the scattering of our molecule
from a potential, to first order in the potential.
If the initial state of the molecule is $|{\mathbf k},l\rangle$,
and we want to find the amplitude to scatter into the state
$|{\mathbf k}^{\prime},l^{\prime}\rangle$.  To lowest order in the
potential, the S-matrix element for this process is
\begin{eqnarray}
\label{smatrix}
\langle{\mathbf k}^{\prime},l^{\prime}|S|{\mathbf k},l\rangle & = &
\delta^{(2)}({\mathbf k}^{\prime}-{\mathbf k})\delta_{l,l^{\prime}}
-2\pi i \delta\left( \frac{(k^{\prime})^{2}}{4m}+\frac{(l^{\prime})^{2}}
{2I}-\frac{k^{2}}{4m}-\frac{l^{2}}{2I}\right)
\langle{\mathbf k}^{\prime},l^{\prime}|V|{\mathbf k},l\rangle .
\end{eqnarray}
In order to find the scattering amplitude for a more general initial state,
\begin{equation}
|\Psi_{in}\rangle = \sum_{l=-\infty}^{\infty}\int d^{2}k \Psi_{in}({\mathbf k},
l)|{\mathbf k},l\rangle ,
\end{equation}
we simply multiply both sides of Eq.\ (\ref{smatrix}) by $\Psi_{in}
({\mathbf k},l)$, integrate over ${\mathbf k}$, and
sum over $l$.

In order to find the cross section, we shall follow the treatment in
Reference \cite{wichmann}. The scattered wave function, $\Phi_{s}
({\mathbf k},l)$ is given by
\begin{equation}
\label{phis}
\Phi_{s}({\mathbf k}^{\prime},l^{\prime})  =  -2\pi i\sum_{l=-\infty}^{\infty}
\int d^{2}k
\delta\left( \frac{(k^{\prime})^{2}}{4m}+\frac{(l^{\prime})^{2}}{2I}
-\frac{k^{2}}{4m}-\frac{l^{2}}{2I}\right)
\langle{\mathbf k}^{\prime},l^{\prime}|V|{\mathbf k},l\rangle
\Psi_{in}({\mathbf k},l)  .
\end{equation}
If $\hat{\mathbf{u}}$ is a unit vector in the ${\mathbf k}^{\prime}$
direction, then the probability of the particle scattering in the
$\hat{\mathbf{u}}$  direction, $P(\hat{\mathbf{u}})$ is
\begin{equation}
P(\hat{\mathbf{u}}) = \sum_{l^{\prime}=-\infty}^{\infty}\int dk^{\prime}\,
k^{\prime}|\Phi_{s}(k^{\prime}\hat{\mathbf{u}},l^{\prime})|^{2} .
\end{equation}
Substituting Eq.\ (\ref{phis}) into the above equation and evaluating the
$k^{\prime}$ integral gives us
\begin{eqnarray}
\label{p(u)}
P(\hat{\mathbf{u}}) & = & 8\pi^{2}m\sum_{l^{\prime}=-\infty}^{\infty}
\sum_{l_{1},l_{2}=-\infty}^{\infty}\int d^{2}k_{1}\int d^{2}k_{2}
\delta\left( \frac{k_{1}^{2}}{4m}+\frac{l_{1}^{2}}{2I}
-\frac{k_{2}^{2}}{4m}-\frac{l_{2}^{2}}{2I}\right) \nonumber \\
 &\times & \Theta\left(\frac{k_{1}^{2}}{4m}+\frac{l_{1}^{2}}{2I}
-\frac{(l^{\prime})^{2}}{2I}\right)
\langle{\mathbf k}_{2},l_{2}|V|\kappa (k_{1},l_{1};l^{\prime})\hat{
\mathbf{u}},l^{\prime}\rangle
 \langle\kappa (k_{1},l_{1};l^{\prime})\hat{
\mathbf{u}},l^{\prime}|V|{\mathbf k}_{1},l_{1}\rangle \Psi_{in}^{\ast}
({\mathbf k}_{2},l_{2})\Psi_{in}({\mathbf k}_{1},l_{1})  ,
\end{eqnarray}
where $\kappa$ is defined in Eq.\ (\ref{kappa}).

Now let us consider an incoming beam of particles that scatter off of
the potential.  The particles are in wave packets $\Phi_{in}({\mathbf k},l)
=\phi_{in}({\mathbf k})\psi_{l}$, where $\phi_{in}$ is highly localized
about ${\mathbf k}=k\hat{{\mathbf y}}$.  The incoming beam is of width
$2R$, so that the center of the wave packets can be displaced in the
$x$ direction anywhere between $-R\leq x \leq R$.  Therefore, we
consider incoming wave functions of the form
\begin{equation}
\Psi_{in}({\mathbf k},l)=e^{-ik_{x}x}\Phi_{in}({\mathbf k},l) ,
\end{equation}
where $-R\leq x \leq R$.  Consequently, we replace $\Psi_{in}({\mathbf k},l)$
in Eq.\ (\ref{p(u)}) by the above expression, and then average the
result over $x$, i.e.\ we find $(1/2R)\int_{-R}^{R}dx P(\hat{\mathbf{u}})$.
Now $2RP(\hat{\mathbf{u}})$ is just the length of the part of the incoming
beam that scatters in the direction $\hat{\mathbf{u}}$, and this is just
the cross section, $\sigma (\hat{\mathbf{u}})$.  We, therefore, have
\begin{equation}
\sigma (\hat{\mathbf{u}}) = \int_{-R}^{R}dx P(\hat{\mathbf{u}}) .
\end{equation}
Assuming the scattering center is much smaller than the beam width, we can
take the limit $R\rightarrow\infty$ in this equation, which introduces
a $2\pi\delta (k_{1x}-k_{2x})$ into the integrals appearing in
Eq.\ (\ref{p(u)}).  The result is
\begin{eqnarray}
\sigma (\hat{\mathbf{u}})& = & 16\pi^{3}m\sum_{l^{\prime}=-\infty}^{\infty}
\sum_{l_{1},l_{2}=-\infty}^{\infty}\int d^{2}k_{1}\int d^{2}k_{2}
\delta\left( \frac{k_{1}^{2}}{4m}+\frac{l_{1}^{2}}{2I}
-\frac{k_{2}^{2}}{4m}-\frac{l_{2}^{2}}{2I}\right)
\delta (k_{1x}-k_{2x}) F({\mathbf k}_{1},{\mathbf k}_{2};l^{\prime},
l_{1},l_{2}) ,
\end{eqnarray}
where
\begin{eqnarray}
F({\mathbf k}_{1},{\mathbf k}_{2};l^{\prime},l_{1},l_{2}) & = & \Theta\left(
\frac{k_{1}^{2}}{4m}+\frac{l_{1}^{2}}{2I}-\frac{(l^{\prime})^{2}}{2I}\right)
\langle{\mathbf k}_{2},l_{2}|V|\kappa (k_{1},l_{1};l^{\prime})\hat{
\mathbf{u}},l^{\prime}\rangle  \nonumber \\
 &\times & \langle\kappa (k_{1},l_{1};l^{\prime})\hat{\mathbf{u}},l^{\prime}|V|
{\mathbf k}_{1},l_{1}\rangle \phi_{in}^{\ast}({\mathbf k}_{2})
\phi_{in}({\mathbf k}_{1})\psi_{l_{2}}^{\ast}\psi_{l_{1}} .
\end{eqnarray}

Our remaining task is to evaluate the integrals.  Let us first do the
${\mathbf k}_{2}$ integral.  The integral over $k_{2x}$ simply sets
$k_{2x}=k_{1x}$, and then the $k_{2y}$ integral sets $k_{2y}=\pm
\kappa (k_{1y},l_{1};l_{2})$.  The result is
\begin{eqnarray}
\sigma (\hat{\mathbf{u}})& = & 16\pi^{3}m\sum_{l^{\prime}=-\infty}^{\infty}
\sum_{l_{1},l_{2}=-\infty}^{\infty}\int d^{2}k_{1}\frac{2m}{\kappa (k_{1y},
l_{1};l_{2})}\sum_{\pm} F(k_{1x},k_{1y},k_{1x},\pm\kappa (k_{1y},l_{1};l_{2}); l^{\prime},l_{1},
l_{2} ) .
\end{eqnarray}
Now suppose that $\phi_{in}({\mathbf k})=\phi_{inx}(k_{x})\phi_{iny}(k_{y})$,
where $\phi_{inx}$ is localized about zero and $\phi_{iny}$ is localized
about $k$.  Then the effect of doing the $k_{1x}$ integral is simply to
set $k_{1x}=0$ in the integrand.  The result is
\begin{eqnarray}
\sigma (\hat{\mathbf{u}}) & = &16\pi^{3}m\sum_{l^{\prime}=-\infty}^{\infty}
\sum_{l_{1},l_{2}=-\infty}^{\infty}\int dk_{1y}\frac{2m}{\kappa (k_{1y},
l_{1};l_{2})}\Theta\left(\frac{k_{1y}^{2}}{4m}
+\frac{l_{1}^{2}}{2I}-\frac{(l^{\prime})^{2}}{2I}\right)  \nonumber \\
 & & \sum_{\pm}\langle\pm\kappa (k_{1y},l_{1};l_{2})\hat{\mathbf{y}}, l_{2}|
V|\kappa (k_{1y},l_{1};l^{\prime})\hat{\mathbf{u}},l^{\prime}\rangle
\langle\kappa (k_{1y},l_{1};l^{\prime})\hat{\mathbf{u}},l^{\prime}|
V|k_{1y}\hat{\mathbf{y}},l_{1}\rangle \phi_{iny}^{\ast}(\pm\kappa
(k_{1y},l_{1};l_{2}))\phi_{iny}(k_{1y}) \psi_{l_{2}}^{\ast}\psi_{l_{1}}
\end{eqnarray}
We only get a substantial contribution to the remaining integral when
$k_{1y}\sim k$ and $\kappa (k_{1y},l_{1};l_{2})\sim k$.  Note that this
implies that only the $+$ in the sum over $+$ and $-$ contributes.  These
two conditions imply that $\kappa (k,l_{1};l_{2})\sim k$, which further
implies that $l_{1}=l_{2}$.  Doing the $k_{1y}$ integral, then, has the
effect of setting $k_{1y}= k$ and $l_{1}=l_{2}$ in the integrand, and
the result is Eq.\ (\ref{gencrosssec}).

%\end{multicols}
\end{document}